\documentclass[pdflatex, sn-chicago, iicol]{sn-jnl}

\usepackage{graphicx}%
\usepackage{multirow}%
\usepackage{amsmath,amssymb,amsfonts}%
\usepackage{amsthm}%
\usepackage{mathrsfs}%
\usepackage[title]{appendix}%
\usepackage{xcolor}%
\usepackage{textcomp}%
\usepackage{manyfoot}%
\usepackage{booktabs}%
\usepackage{algorithm}%
\usepackage{algorithmicx}%
\usepackage{algpseudocode}%
\usepackage{listings}%

\usepackage{import}
\usepackage{mathtools}
\usepackage{dsfont,bm}
\usepackage{cleveref}
\usepackage{algorithm,algpseudocode}

\graphicspath{{./figures}}

\newcommand*\xbar[1]{%
	\hbox{%
		\vbox{%
			\hrule height 0.5pt 
			\kern0.5ex
			\hbox{%
				\kern-0.1em
				\ensuremath{#1}%
				\kern-0.1em
			}%
		}%
	}%
} 

\newcommand{\slfrac}[2]{\left.#1\middle/#2\right.}

\newcommand{\beq}{\begin{equation}}
\newcommand{\eeq}{\end{equation}}
\newcommand{\bal}{\begin{align}}
\newcommand{\eeal}{\end{align}}
\newcommand{\baln}{\begin{align*}}
\newcommand{\ealn}{\end{align*}}

\newcommand{\norm}[1]{\lVert#1\rVert}

\renewcommand \v {\bm} 

\newcommand{\nablad}{{\nabla}}                        %

\newcommand{\tp}{\intercal} 
\newcommand{\eref}[1]{(\ref{#1})}
\newcommand{\sref}[1]{section~\ref{#1}}
\newcommand{\fref}[1]{figure~\ref{#1}}
\newcommand{\Eref}[1]{equation (\ref{#1})}

\newcommand{\bcite}[1]{(\cite{#1})}

\newcommand{\I}[1][]{\mathcal{J}_{#1}}           
\newcommand{\f}[1][]{\widetilde{\mathcal{F}}}           
\newcommand{\fq}[1][]{\widetilde{\mathcal{F}}_{\q}}           
\newcommand{\fcv}[1][]{\widetilde{\mathcal{F}}_{\cv}}           
\newcommand{\fS}[1][]{f}           
\newcommand{\fSq}[1][]{f_{\q}}           

\newcommand{\La}{\mathcal{L}}           
\newcommand{\Laug}[1][]{L^a_{#1}}           

\newcommand{\Ir}[1][]{\hat{\mathcal{J}}_{#1}}
\newcommand{\gr}[1][]{\bm{\hat{g}_{#1}}}
\newcommand{\gri}[1][]{{\hat{g}_{i}}}

\newcommand{\Lar}[1][]{\hat{\mathcal{\La}}}

\newcommand{\Ompe}{\Omega_{\textrm{pe}}}  
\newcommand{\IOmpe}[1][]{\mathds{1}_{\Ompe}} 
\newcommand{\Omc}{\Omega_{\textrm{c},{i}}}  
\newcommand{\IOmc}[1][]{\mathds{1}_{\Omc}} 
\newcommand{\cv}[1][]{{\Omega}_{#1}}
\newcommand{\SV}[1][]{\boldsymbol{\mathcal{V}}_{#1}}

\newcommand{\p}{\v{r}}                 
\newcommand{\pu}{\v{r_{uc}}}                 
\newcommand{\pij}{\v{r_{i,j}}}                 

\newcommand{\xu}{{x_{uc}}}                 
\newcommand{\xij}{{x_{i,j}}}                 

\newcommand{\tr}{\delta_{\textrm{tr},k}}

\newcommand{\om}[1][]{\boldsymbol{\omega}_{#1}}              

\newcommand{\Omf}{{\Omega_{}}}

\newcommand{\ex}{{\bf e}_{x}} 

\newcommand{\Nufs}{{\boldsymbol \nu}_{f}}    

\newcommand{\dd}{\mathrm{d}}                                
\newcommand{\av}[2][]{\left\langle #2 \right\rangle_{#1}}

\newcommand{\dfs}{\delta_{fs}}

\newcommand{\ddS}{\dd \sigma}

\newcommand{\intsb}[2]{\int_{\Sigma_{\mathrm{#1}}} #2 \,\ddS}

\newcommand{\q}[1][]{{\boldsymbol x}_{#1}}     
\newcommand{\y}[1][]{{\boldsymbol y}_{#1}}               

\newcommand{\ud}[1][]{{\boldsymbol u}_{#1}}
\newcommand{\stress}[1][]{{\boldsymbol \tau}_{#1}}

\newcommand{\pf}[1][]{ {p^{*}}_{#1}}
\newcommand{\pfd}[1][]{ {p^{*}}_{#1}}
\newcommand{\Pav}[1][]{{ P}_{#1}}

\newcommand{\dPavd}[1][]{{\nabla P}_{#1}}
\newcommand{\dPavx}[1][]{{\nabla_x P}_{#1}}

\newcommand{\Qtot}{{Q}_\textrm{unit}}
\newcommand{\Qtotb}{{Q}_\textrm{unit,s}}
\newcommand{\Qa}{{Q}_\textrm{array}}
\newcommand{\Qp}{{Q}_\textrm{periodic}}

\newcommand{\Qij}{{Q}_{i,j}}

\newcommand{\z}[1][]{{\boldsymbol z}_{#1}}               

\newcommand{\Tin}{T_\textrm{in}}                 
\newcommand{\Tout}{T_\textrm{b,out}}                 
\newcommand{\To}{T_0}                 
\newcommand{\Tob}{T_{1}}                 
\newcommand{\Ts}{T_s}                 
\newcommand{\Td}{\theta}                 
\newcommand{\lT}{\lambda_T}				 
\newcommand{\lTd}{{\lambda}_T}		 

\newcommand\Rey{\mbox{\textit{Re}}}  
\newcommand\Pran{\mbox{\textit{Pr}}} 

\newcommand{\E}{\v{E}} 
\renewcommand{\H}{\v{H}} 
\newcommand{\Hp}{\v{H}_{\dPavd}} 
\newcommand{\Hu}{\v{H}_{u_x}} 

\theoremstyle{thmstyleone}%
\theoremstyle{thmstyletwo}%

\theoremstyle{thmstylethree}%

\begin{document}

\title[Article Title]{A Unit-Cell Shape Optimization Approach for Maximizing Heat Transfer in Periodic Fin Arrays at Constant Solid Temperature}


\author*[1,2]{\fnm{Maarten} \sur{Blommaert}}\email{maarten.blommaert@kuleuven.be}

\author[1,2]{\fnm{Arthur} \sur{Vangeffelen}}
\author[1,2]{\fnm{Mehmet} \sur{Basaran}}
\author[2,3]{\fnm{Geert} \sur{Buckinx}}
\author[1,2]{\fnm{Martine} \sur{Baelmans}}



\affil*[1]{\orgdiv{Department of Mechanical Engineering}, \orgname{KU Leuven}, \orgaddress{\street{Celestijnenlaan 300}, \postcode{3001} \city{Leuven}, \country{Belgium}}}

\affil[2]{\orgname{EnergyVille}, \orgaddress{\street{Thor Park}, \postcode{3600} \city{Genk}, \country{Belgium}}}

\affil[3]{\orgname{Flemish Institute for Technological Research (VITO)}, \orgaddress{\street{Boeretang 200}, \postcode{2400} \city{Mol}, \country{Belgium}}}


\abstract{Periodic fin structures are often employed to enhance heat transfer in compact cooling solutions and heat exchangers. Adjoint-based optimization methods are able to further increase the heat transfer by optimizing the fin geometry. However, obtaining optimal geometries remains challenging in general because of the high computational cost of full array simulations. In this paper, a unit cell optimization approach is presented that starts from recently developed macro-scale models for isothermal solid structures.  The models exploit the periodicity of the problem to reduce the computational cost of evaluating the array heat transfer to that of a single periodic unit cell. By combining these models with a geometrically-constrained free-shape optimization approach, optimal fin geometries are obtained for the periodic fin array that maintain a minimal fin distance. Moreover, using an augmented Lagrangian approach, also the average pressure gradient and barycenter of the fin can be fixed. On a fictitious use-case, heat transfer increases up to 104 \% are obtained. When also flow rate is constrained in addition to maintain a high effectiveness, only up to 8 \% heat transfer increase is observed. Finally, the errors of the unit-cell optimization approach are investigated, indicating that with a good choice of cost functional formulation, errors of the approach as low as 1-2 \% can be obtained for the periodically developed part of the array. Finally, the entrance effect to the heat transfer is found to be non-negligible with a contribution of 10-15 \% for the considered fin array. This advocates for further research to extend the unit-cell models towards improved modeling of entrance effects.   }

\keywords{unit-cell heat transfer model, free-shape optimization, Augmented Lagrangian method, extended heat transfer surfaces}



\maketitle  
\section{Introduction}
The ongoing miniaturization of electronics necessitates compact heat sinks to guarantee thermal management within confined spaces. Similarly, the role of compact heat exchangers in increasing the energy-efficiency of systems cannot be overstated. Across diverse applications, limitations in space, material, and weight, increase the demand for compact heat transfer devices. Common practice in the design of these devices is to improve heat transfer by the use of periodic fin structures, such as pin fins, wavy fins, serrated fins, or offset-strip fins. 

To further increase heat transfer with respect to these standard fin types, shape or topology optimization methods can be employed. While shape optimization is a well-established practice in, for instance, the aerodynamic design of wing profiles, shape optimization of extended heat transfer surfaces is much less explored in literature. Although the aligned meshes in shape optimization are a complexity to deal with, the approach has clear advantages over density-based topology optimization when it comes to accurately resolving the solid-fluid interfaces. In the literature, periodic fin configurations \bcite{Micheli2008,Cavazzuti2008,Copiello2009} or tube bundles \bcite{Ranut2014} are most commonly represented by a small number of parameters and optimized with genetic algorithms. Only a limited number of studies consider adjoint-based optimization, allowing first demonstrations of free-shape optimization of 3D fin shapes \bcite{Wang2022,Wang2024} or level-set topology optimization of heat transfer surfaces with aligned meshes \bcite{Feppon2021}, showing potential that could be exploited by additive manufacturing methods. At current, the high computational cost of a full fin-shape simulation makes a full simulation of relevant compact heat exchanger configurations intractable. Therefore, these aforementioned 3D optimizations are currently restricted to relatively crude configurations or necessitate ad-hoc model approximations.

To reduce the computational cost of simulating flow and heat transfer through periodic structures in a systematic way, two main theoretical frameworks can be distinguished to characterize the heat transfer regime using unit-cell simulations containing only a single fin. 
The first framework assumes a periodically developed heat transfer regime, where the temperature distribution can be determined by solving periodic flow and heat transfer equations on a single unit cell \bcite{patankar1977fully,buckinx2015isothermal}. 
The second framework employs the volume-averaging technique (VAT), typically used for porous media, to derive a macroscale temperature field. Here, the interfacial heat transfer rate is modeled through a heat transfer coefficient, which is obtained from a closure problem that accounts for local deviations from the macroscale temperature field \bcite{whitaker1996forchheimer,quintard1997two}. 
As demonstrated by Buckinx and Baelmans \bcite{buckinx2015multi,buckinx2015isothermal,buckinx2016macro,buckinx2017macro}, extending the weighted averaging method of Quintard and Whitaker \bcite{quintard1994transport1,quintard1994transport2,quintard1994transport3,quintard1994transport4,quintard1997two,davit2017technical}, both frameworks are theoretically equivalent when a double volume-averaging operation is used for defining the macroscale variables. 
The resulting macroscale model provides an exact formulation for periodically developed flow and heat transfer in periodic fin arrays, both for isothermal solids \bcite{buckinx2015isothermal} and for conjugate heat transfer with constant imposed heat flux \bcite{buckinx2016macro}. In other work, they have shown that flow and heat transfer reach a (quasi-)developed state following a short development region, typically spanning 5 to 15 fin rows \bcite{buckinx2022, buckinx2023, Vangeffelen2023, Vangeffelen2025}. 
 
  This paper proposes a shape optimization framework for periodic fin arrays that takes advantage of these unit-cell models. Specifically, it focuses on the free-shape optimization of periodic fin arrays that are best modeled as isothermal solids. This is for example the case for highly unbalanced heat exchangers, heat exchangers with a phase-changing process at one side, or heat sinks with a highly conductive and sufficiently thick base plate. 
The paper demonstrates how unit-cell models for flow through isothermal solids can be exploited for the free-shape optimization of an array of cylindrical pin fins. To this end, a methodology is elaborated for periodic heat transfer optimization using efficient adjoint shape calculus. The paper therefore aims to:
\begin{enumerate}
    \item present a globalized optimization methodology for extended heat transfer surfaces based on a unit-cell model for an isothermal fin array that allows including relevant geometric and operational constraints;
    \item demonstrate the approach on a use-case and Quantify the potential for heat transfer improvement;
    \item assess the impact of unit-cell modeling errors on the optimization outcome and investigate alternative unit-cell heat transfer formulations that limit the error on the predicted heat transfer increase.
\end{enumerate}

The paper is organized as follows. Section \ref{sec:model} introduces the unit-cell equations, an approach to solving them, and expressions for the array heat transfer. Next, we outline the optimization approach in \sref{sec:opt}. Finally, in \sref{sec:results}, we discuss the optimization of a 2D fin array, investigating the heat transfer increase and unit-cell errors for alternative optimization problem formulations and unit-cell heat transfer formulations.

\section{A unit-cell model for periodically developed flow and heat transfer}\label{sec:model}
\begin{figure}
	\centering
	\includegraphics[width=1.0\columnwidth]{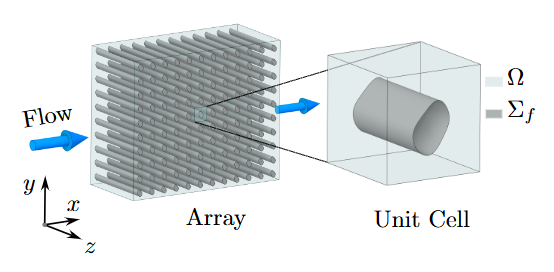}
	\caption[]{ This figure illustrates a periodic fin array, highlighting a unit cell and its fluid domain $\cv$ and fin boundary $\Sigma_f$. }
	\label{fig:array}
\end{figure}

The unit cell shape optimization approach proposed in this paper aims to optimize the shape of an isothermal solid fin that is periodically repeated in a fin array (see \fref{fig:array}). To reduce the computational cost of a full flow and heat transfer simulation of the fin array, the periodicity of the fin configuration, flow, and heat transfer can be exploited. To this end, the unit cell modeling approach of \cite{buckinx2015isothermal} is used. Below, it is briefly summarized for consistency in \sref{sec:21}. Subsequently, in \sref{sec:22}, we describe the array heat transfer in terms of unit-cell quantities.  

\subsection{Model equations}\label{sec:21}
By assuming the flow and heat transfer are periodically developed throughout, the model decomposes the governing equations into detailed microscale fluctuations on the unit-cell level and overall macroscale evolutions on the array level.  While the fluid velocity $\ud$ only varies on the unit-cell level, the pressure $p$ decomposes into a linear overall evolution on the array (macroscale) level and a periodic variation on the unit-cell (microscale) level $\pf$, i.e. 
\begin{equation}\label{eq:01} p = \nabla \Pav \cdot ( \p - \p_0) + \pf, \end{equation}  
with $\nabla \Pav$ a constant overall pressure gradient and $\p$ and $\p_0$ denote position vectors. 
Assuming steady incompressible flow, the periodic contributions to flow and pressure are then governed by
\begin{equation}\label{eq:02} 
	\begin{aligned}
		&\nablad \cdot \ud = 0 \\
		&\rho \ud \cdot \nablad \ud =  - \nablad \pfd - \dPavd +\nablad \cdot \stress,
	\end{aligned} 
\end{equation}
with $\rho$ the fluid density and $\stress= \mu \nabla \ud + \mu \left(\nabla \ud\right)^\tp$ the shear stress tensor.
These equations are augmented with no-slip boundary conditions at fin-fluid interfaces,  and periodic boundary conditions at the unit-cell edges for $\ud$ and $\pfd$. In the flow direction, the mean pressure gradient can either be imposed directly, or determined to match an imposed mean velocity. Assuming the mean flow is oriented in the $x$-direction, this gives
\begin{subequations}
	\begin{align}\label{eq:03a} 
		&\dPavx = \left[{\nabla_x \Pav}\,\right]_\textrm{fix}\textrm{ or}\\\label{eq:03b} 
		&\av{u_x} = u_{x,\textrm{fix}},  
	\end{align} 
\end{subequations}
respectively. $\nabla_x$ hereby denotes the derivative in the x-direction. The notation $\av{\phi}$ refers to the spatial average over the unit cell, evaluated here as
\begin{equation}
\av{\phi} = \frac{\varepsilon}{V_f}\int_\Omf \phi \ d\omega,
\end{equation}
with $\Omf$ the fluid domain in the unit cell (see \fref{fig:array}) with volume $V_f$, and $\varepsilon$ the porosity of the unit cell.
In the $y$-direction, we choose here to maintain no net flow to physically respect the closed sides of the device, i.e. 
\begin{equation} \label{eq:04} 
	\av{u_y} = 0,  
\end{equation}
which can lead to a nonzero vertical pressure gradient for asymmetric shapes. The flow equations, eqs. \eref{eq:01} to \eref{eq:04}, will further succinctly be denoted as $\H(\cv, \y) = 0$, with subscript $u$ or $\dPavd$ depending on whether condition \eref{eq:03a} or \eref{eq:03b} is applied, respectively. $\y = [\ud^\tp,\pf]^\tp$ hereby represents the vector of flow state variables.


To model the heat transfer based on a unit-cell approach, the temperature field \( T \) is decomposed into an exponentially decaying macroscale field with decay rate \( \lT \) and a  contribution \( \Td \) that is spatially periodic over each unit cell:  
\begin{align}
\label{eq:06}
T - \Ts &= \To \frac{\Td}{\av{\Td}} \exp\left(\lT x\right),
\end{align}
Here, \( \To \) is an unknown constant that sets the absolute temperature at the onset of the developed heat transfer regime.
Alternatively, this can be written as
\begin{align}
\label{eq:06b}
T -\Ts&=  \Tob \frac{\Td}{\norm{\Td}_\infty} \exp\left(\lT x\right),  
\end{align}
where $\Tob$ is a rescaling of $\To$, namely 
\begin{equation}
    \Tob = \To \frac{\norm{\Td}_\infty}{\av{\Td}}.
\end{equation} The benefit of renormalizing the temperature mode \( \Td \) in this way for optimization purposes will be discussed later.

The steady energy equation then decomposes into a detailed transport model for the periodic contribution to temperature in a single unit cell $\Td$ and an eigenvalue equation for the macroscale temperature decay rate $\lT$,
\begin{subequations}
	\begin{align}
		&\label{eq:micro1} \rho \nabla \cdot \left(\ud c\Td\right) = \nablad \cdot \left(k\nablad \Td \right)+ \sigma_T\\
		&\label{eq:micro2}\sigma_T = \left(2 k\nablad \Td - \rho c \ud \Td \right) \cdot \ex \lTd + k \Td \lTd^2\\
		&\label{eq:micro3}\av{\Nufs\cdot k\nablad \Td \, \dfs} - \rho c \left(\av{\ud\Td}\cdot \ex\right) \lTd +k\av{\Td} \lTd^2 = 0.
	\end{align}
\end{subequations}
These equations are supplemented with periodic boundary conditions for $\Td$ on the boundaries of the unit cells and $\Td=0$ at the edge of the fin. Moreover, a scalar condition is imposed to fix the scaling of $\Td$. In this paper, we choose to impose a Dirichlet condition, $\Td = 1$, in a corner of the unit cell. 

The quadratic eigenvalue equation \eref{eq:micro1} can also be written explicitly in terms of its negative root, 
\begin{equation}\label{eq:08}
	\lTd =  \frac{\av{\ud\Td}\cdot \ex}{2\alpha{\av{\Td}}}  - 
	\sqrt{\left(\frac{\av{\ud\Td}\cdot \ex}{2\alpha\av{\Td}}\right)^2- \frac{\av{\Nufs\cdot \nablad \Td \, \dfs}}{{\av{\Td}}}},
\end{equation}
since only the negative root provides a physically sensible solution. To avoid nonphysical solutions emerging during optimization, equation \eref{eq:08} is used to solve directly for the negative root.  
The governing energy equations \eref{eq:micro1}, \eref{eq:micro2}, \eref{eq:08}, and their corresponding boundary conditions can then succinctly be denoted as $\E(\cv, \y,\z) = 0$, with $\z = [\Td,\lTd]^\tp$ the thermal state variables.


\subsection{Solution of the model equations}\label{sec:22a}
The model presented in \sref{sec:21} comprises a closed set of unit-cell model equations that can be solved for the unit-cell state vectors $\y$ and $\z$. 
They show a clear sequential structure that is exploited in their solution. The simulation is started from a given fin shape that is implicitly parameterized by the unit-cell fluid domain $\cv$ that it delimits. Starting from this fin shape, the flow equations $\H(\cv, \y) = 0$ are solved using a coupled solver for the flow state variables $\y$. Then, the periodic energy equations $\E(\cv, \y,\z) = 0$ are solved for the thermal state variables $\z$. 

The weak form of these equations is hereto discretized and solved using the FEniCS finite element library (\cite{fenics2010,fenics2012}), using Taylor-Hood elements for the periodic flow equations and second-order Lagrangian elements for the periodic energy equation. Newton-based iterative solvers are then used to solve the nonlinear algebraic equations. 

\subsection{Array heat transfer}\label{sec:22}
Starting from the model presented in \sref{sec:21}, the heat transfer rate across the fluid-solid interface in the complete fin array can now be written in terms of the unit-cell state variables $\y$ and $\z$. 
  
In a periodic fin array with $n_x$ fins in the streamswise direction and $n_y$ fins in the transverse direction, the rate of heat $\Qij$ exchanged on the surface $\Sigma_\textrm{f}$ of a fin in unit cell $i,j$ is
\begin{equation}
	\begin{aligned}
		\Qij &= \intsb{f}{k \nabla T \cdot\Nufs},
	\end{aligned}
\end{equation}
with ${\Nufs}$ the unit normal on the fin surface $\Sigma_f$ pointing inwards in the fluid domain.
To find an expression as a function of the periodic model variables, we
insert the decomposition of the temperature according to \eref{eq:06}, leading to 
\begin{equation}
	\begin{aligned}
		\Qij &= \intsb{f}{k\To \frac{\nabla \Td}{\av{\Td}}\exp\left(\lT x\right) \cdot\Nufs}
	\end{aligned}.
\end{equation}
 
 One can relate the streamwise position $\p$ of the array reference frame to the position $\pu$ in a relative frame with its origin fixed to the center of a unit cell $i,j$ with position $\pij$.  This gives
\begin{equation}
	\p = \pij+\pu.
\end{equation}
Taking only the streamwise component analogously gives
\begin{equation}
	{x} = \xij +\xu.
\end{equation}
This allows rewriting the exchanged heat in this cell as
\begin{equation}
	\begin{aligned}
		\Qij &= \To \exp\left(\lT \xij\right)\intsb{f}{k \frac{\nabla \Td}{\av{\Td}}\exp\left(\lT \xu\right) \cdot\Nufs}
	\end{aligned}.
\end{equation}
The heat transfer of the full array then becomes
\begin{subequations}\label{eq:cost}
	\begin{align}
		 \label{eq:cost1}&\Qtot(\cv,\z) \\\nonumber&=  k \To\sum_{i,j}\exp\left(\xij\right)\intsb{f}{ \frac{\nabla \Td}{\av{\Td}}\exp\left(\lTd \xu\right) \cdot\Nufs} \intertext{or}
		\label{eq:cost2}&\Qtotb(\cv, \z) \\\nonumber& =  k \Tob\sum_{i,j}\exp\left(\xij\right)\intsb{f}{ \frac{\nabla \Td}{\norm{\Td}_\infty}\exp\left(\lTd \xu\right) \cdot\Nufs},
	\end{align}
\end{subequations}
with the alternative temperature decomposition of \Eref{eq:06b}. Equation \eref{eq:cost1} will further be referenced to as the ``normalized" heat transfer formulation and equation \eref{eq:cost2} as the ``scaled" formulation.  It should be noted that the scale factors $\To$ and $\Tob$ depend on the fin shape and flow conditions in the fin array. As such, they are not merely constant factors but a function of the design variables $\cv$.  

\section{Optimization approach}\label{sec:opt}
In this section, an optimization approach is presented that allows optimizing the shape of the fins in a periodic fin array at constant solid temperature. 
\subsection{Optimization problem formulation}
\begin{figure}
	\centering
	\includegraphics[width=0.9\columnwidth]{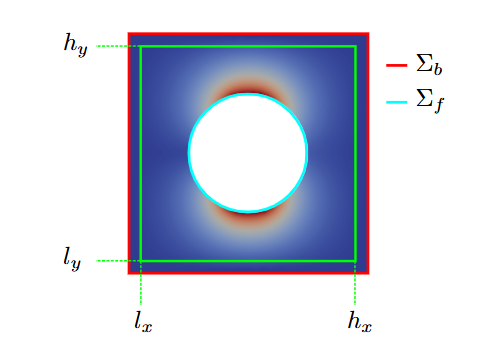}
	\caption[]{Figure illustrating the components of the shape optimization subproblem, including the geometrical box to which the fin is constrained and the unit-cell domain boundaries. The colors illustrate the magnitude of the initial domain deformation field $\SV$.   
	}%
	\label{fig:fin}
\end{figure}

The shape of the fluid domain in the unit cell $\cv$ is hereby optimized to maximize the heat transfer of the fin array. If a constant pressure drop needs to be maintained over the array, the macroscopic pressure gradient $\dPavd$ will be fixed in the x-direction. The optimization problem can then be formulated as   
\begin{equation} \label{eq:opt}\begin{aligned} \min_{\Omega \in \mathcal{O},\y,\z}& \qquad
\I \left( \cv,\z\right) = -\Qtot\left( \cv,\z\right), \\
s.t.& \qquad \Hp(\cv, \y) = 0, \\
& \qquad \E(\cv, \y,\z) = 0,\\
& \qquad \int_\Omega{\om}\ {\textrm{d}}\om = 0.
\end{aligned}\end{equation}
The last constraint hereby fixes the barycenter of the fin to the center of the unit cell, with $\om \in \Omega$ the location of a point in the unit-cell fluid domain. This serves as a regularization for the periodic optimization problem. 
The feasible set of fluid domains, $\mathcal{O}$, implicitly imposes geometric restrictions on the fin shape to prevent topological changes, ensuring it does not approach the unit-cell boundary. To enforce this, the fin is confined within a geometric box, as shown in \fref{fig:fin}.

It should be noted that in the heat transfer objective \eqref{eq:cost} the scaling variables $\To$ and $\Tob$ cannot be determined from the unit-cell equations. To determine the absolute value of the heat transfer, a direct simulation of the complete array is thus needed. Throughout the optimization, these scaling variables will be deliberately assumed constant so that they do not need to be accounted for.  
Moreover, by introducing the barycenter and geometric constraints, the maximum value of \(\Td\) systematically occurs at the domain corners, yielding \(\|\Td\|_\infty = 1\) in the cost functional formulation of \eqref{eq:cost2}. This corresponds to the Dirichlet condition \(\Td = 1\) imposed at the unit-cell domain corners.

As will be discussed in the results section, constraining only the pressure drop over a unit cell creates an incentive for the optimizer to increase the heat transfer by simply increasing the flow rate through a reduction in hydraulic resistance over the fin array. In heat exchangers or heat sinks where pumping costs are substantial, this may be undesirable. A second optimization formulation will therefore be explored in this paper where both flow rate and pressure are fixed by introducing an additional constraint, i.e.
\begin{equation} \label{eq:16}\begin{aligned} \min_{\Omega \in \mathcal{O},\y,\z}& \qquad
		\I \left( \cv,\z\right) = -\Qtot\left( \cv,\z\right), \\
		s.t.& \qquad \Hu(\cv, \y) = 0, \\
		& \qquad \E(\cv, \y,\z) = 0,\\
		& \qquad \dPavx = \left[{\nabla_x \Pav}\,\right]_\textrm{fix},\\
	&\qquad \int_\Omega{\om}\ {\textrm{d}}\om = 0.
\end{aligned}\end{equation}

\subsection{Solution approach}
Remark that the unit-cell state equations create a direct functional dependence of the state variables on the fin shape $\cv$ which could in short be written as $\y\left(\cv\right)$ and $\z = \left(\cv\right)$. This mapping can be used to eliminate $\y$ and $\z$ from the optimization problem. The reduced version of \Eref{eq:16} thus becomes
\begin{equation} \begin{aligned} \min_{\Omega \in \mathcal{O}}& \qquad
		\Ir \left( \cv\right),  \\
		s.t.
		&  \qquad \gr(\cv) = 0,
\end{aligned}\end{equation}
with $\Ir$ the reduced objective functional. The reduced constraint vector $\gr(\cv)$ hereby groups the additional constraints imposed on the optimization problem. In the following subsection we describe how this reduced optimization problem is solved.
\subsubsection{Augmented Lagrangian solution approach}
The optimization problem is solved using an in-house Augmented Lagrangian solver, based on the algorithm described in \cite{Nocedal2006}. To this end, an Augmented Lagrangian
\begin{equation}
	\Laug \left(\cv, \v{\lambda};\mu\right)= \Ir \left( \cv\right) - \sum_{i=1}^m \lambda_i \gri(\cv) + \frac{\mu}{2}\sum_{i=1}^m \lambda_i \left(\gri(\cv)\right)^2
\end{equation}
is introduced, for which a series of unconstrained optimization problems
\begin{equation} \label{eq:19}\begin{aligned} \min_{\Omega \in \mathcal{O}}& \qquad
		\Laug \left(\cv, \v{\lambda};\mu\right)
\end{aligned}\end{equation}
is solved, with changing variables $\v{\lambda}$ and $\mu$. 

\subsubsection{Quadratic shape optimization subproblems}
To optimize the shape of the fin $\cv$, these subproblems are solved in each Augmented Lagrangian iteration as a sequence of quadratic programming problems for the domain deformation $\SV$ in each iteration $k$, i.e.
\begin{subequations} \label{eq:20}\begin{align} \min_{\SV}& \label{eq:20a} \qquad
		\frac{1}{2\alpha} \SV^\tp \left(I-e \Delta\right)\SV + \left(\nabla \Laug[k]\right)^\tp \SV,\\
		s.t.& \qquad\norm{\SV}_\infty < \tr,\\
		& \label{eq:gc} \qquad l_j \leq \left[\om + \SV \right]_j\leq h_j \quad \om,\SV \in \Sigma_f,
		\\&\label{eq:20d}\qquad \SV=0 \quad \SV \in \Sigma_b.
\end{align}\end{subequations}
By applying the Laplacian smoothing operator $\Delta$ to the shape deformation in the entire fluid domain, a mesh deformation strategy is automatically achieved, with $e$ a numerical parameter to tune the smoothing. In a discrete sense, the vector field $\SV$ gives hereby a position update for all mesh nodes. The parameter $\alpha$ serves as a scaling for the size of the allowed shape deformations in a single iteration. The constraint $\norm{\SV}_\infty < \tr$ limits the shape deformations in a single iteration to the trust region radius $\tr$, which is updated in a trust-region globalization strategy. The constraints in \Eref{eq:gc} limit the $x$- and $y$-components of the fin position to the geometrical box as illustrated in \fref{fig:fin}, with $l_j$ and $h_j$ the lower and upper bounds for each of these two coordinate axes. Equation \eref{eq:20d} finally imposes that the outer domain boundary $\Sigma_b$ should not deform. After discretization, this reduces to a quadratic program (QP) that can be solved with a standard QP solver.

\subsubsection{Shape sensitivities}
The shape sensitivities  $\nabla \Laug[k]$ in the subproblem \eref{eq:20} are obtained using the adjoint method. For the implementation of the adjoint method, this work draws inspiration from the FEniCS-based approaches in \cite{Schmidth2018} and \cite{ham2018}. 
The methodology involves differentiating the Lagrangian in its weak form with respect to the adjoint, state, and design variables (i.e. mesh node coordinates) using the automatic differentiation capabilities of the UFL language, to obtain the state and adjoint equations, and the design derivative. 
Once the state and adjoint equations are solved, the weak shape sensitivities of the Augmented Lagrangian are automatically generated following the approach in \cite{ham2018}.
\subsubsection{Shape optimization algorithm}
The Augmented Lagrangian subproblem of \Eref{eq:19} leads to a series of box-constrained optimization problems. The solution procedure for each such subproblem is given in the algorithm \ref{alg:alg1}. 

\begin{algorithm}
	\caption{Shape optimization algorithm with trust region globalization}
	\label{alg:alg1}
	\begin{algorithmic}[1]
		\State Retrieve $\v{\lambda}$ and $\mu$, and initial shape $\cv^*$ from the Augmented Lagrangian algorithm
		\While{Termination condition not reached}
		\State Perform a periodic simulation for $\cv^*$
		\State Evaluate $\Laug\left(\cv^*, \v{\lambda};\mu\right)$
		\State Evaluate $\rho_\textrm{tr} = \frac{\Laug\left(\cv^*, \v{\lambda};\mu\right)-\Laug\left(\cv, \v{\lambda};\mu\right)}{d\Laug}$
		\State Update $\tr$
		\If{ $\rho_\textrm{tr} > 0.15$}
		\State Update design: $\cv \leftarrow \cv^*$
		\State Remesh every $n$ accepted steps
		\State Solve adjoint periodic simulation
		\State Evaluate design gradient $\nabla \Laug[k]$
		\EndIf
		\State Calculate a new box-constrained shape velocity $\SV$ using \Eref{eq:20} and the predicted decrease in $\Laug$ \\$d\Laug = \frac{1}{2\alpha} \SV^\tp \left(I-e \Delta\right)\SV + \left(\nabla \Laug[k]\right)^\tp \SV$
		\State Update tentative design $\cv^* \leftarrow \cv$ by moving points with $\om^* = \om + \SV$
		\EndWhile
	\end{algorithmic}
\end{algorithm}
 The trust-region globalization in the algorithm ensures a stable convergence to the optimum by evaluating the parameter $\rho_\textrm{tr}$ that compares the observed change in $\Laug$ with the one predicted by the quadratic approximation in \Eref{eq:20a}. Note that for sufficiently small steps, $\rho_\textrm{tr} \approx 1$.
 After a number of smooth mesh deformations, mesh quality may decrease and artefacts such as stretched and bunched cells may arise. Therefore, the algorithm includes a remeshing step every $n$-th accepted optimization step. To this end, GMSH version 3.0.6 (\cite{GMSH}) is called to mesh the fluid domain, using the current fin shape as inner mesh boundary. 

\section{Case setup and unit-cell shape optimization results}\label{sec:results}
In this section, the unit-cell optimization methodology is applied to optimize the shape of a periodic fin structure. First, the case setup is introduced. Then, \sref{sec:42} and \sref{sec:43} present the optimization results for imposed pressure drop and for both imposed pressure drop and flow rate, respectively. Finally, \sref{sec:44} examines the errors in the unit-cell optimization approach in greater detail and discusses directions for future research.

\subsection{Case set-up}\label{sec:41}
\begin{figure*}
	\centering
    \includegraphics[width=0.9\textwidth]{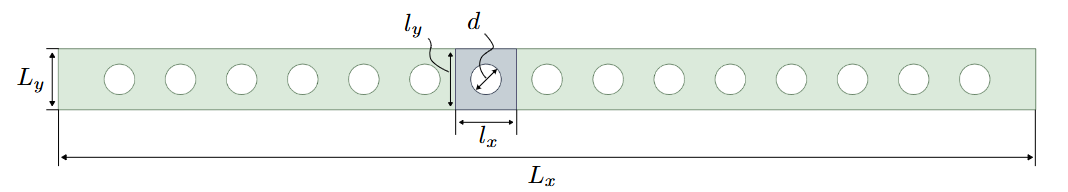}
	\caption[]{Geometry of cylinder array and unit cell. 
    }
	\label{fig:use_case}
\end{figure*}

For simplicity, the optimization strategy of \sref{sec:opt} is applied to the two-dimensional design optimization of an array of fins with constant solid temperature. Physically, the fin array could either represent a fin array of long and highly conductive fins in a heat sink, or a tube array in a cross-flow heat exchanger with a fluid undergoing a phase change in the tubes. Here, we start with a single array of 15 fins with overall dimensions of { $L_x = 16 $ mm, $L_y = l_y = 1$ mm} and a pressure drop \( \Delta p \) of { $12$ Pa} over the array. Cylindrical pin fins with a diameter of \(d = 0.5\) mm and a pitch of \(l_x = 1\) mm, positioned at the center of unit cells measuring \(l_x = 1\) mm by \(l_y = 1\) mm, serve as the reference shape for comparison and as the starting point for optimization, as illustrated in Figure \ref{fig:use_case}. Note that in the lateral ($y$) direction, periodicity is assumed in both unit-cell model and array simulation. The cooling fluid parameters are characteristic for air, with density of { $1.184 \ \slfrac{\textrm{kg}}{\textrm{m}^3}$ }, heat capacity of { $1007 \slfrac{\textrm{J}}{\textrm{kg} \, \textrm{K}}$}, kinematic viscosity of $1.56\times 10^{-5} \, \textrm{m}^2/\textrm{s}$, and a Prandtl number of $\Pran = 1$.

A simulation of this initial array shows a flow rate per unit height of  { $8.08 \times 10^{-7} \ \slfrac{\textrm{kg}}{\textrm{s}\,\textrm{m}}$}, corresponding to a Reynolds number of $\Rey = \slfrac{\av{u_x} l_y }{\nu}= 43$ based on the volume-averaged velocity over the unit-cell, and a total heat transfer per unit height of { $24 \textrm{W} \textrm{m}^{-1}$} from the fins. The temperature of the fluid is shown in \fref{fig:dpopt}a.

\subsection{Fin shape optimization for imposed pressure drop}\label{sec:42}
This cylindrical pin-fin array will now be optimized for maximal heat transfer for a given array pressure drop. To this end, we solve the optimization formulation of \Eref{eq:opt}. 
However, establishing a direct relationship between the imposed periodically developed pressure gradient in \Eref{eq:03a} and the overall pressure drop \( \Delta p \) across the array is non-trivial due to entrance effects. Neglecting the entrance pressure drop can lead to discrepancies in flow rate between the array and the periodic simulation, potentially reducing the relevance of the optimization.  
To address this, we solve \Eref{eq:02} with a prescribed flow rate in \Eref{eq:03b} matching that of the initial array simulation. This approach yields a periodic pressure gradient for the initial array of \(\left[{\nabla_x \Pav}\,\right]_\textrm{fix} = 722.7 \, \text{Pa/m} \), which we enforce throughout the optimization.

\begin{figure*}
	\centering
	\includegraphics[width=0.95\textwidth]{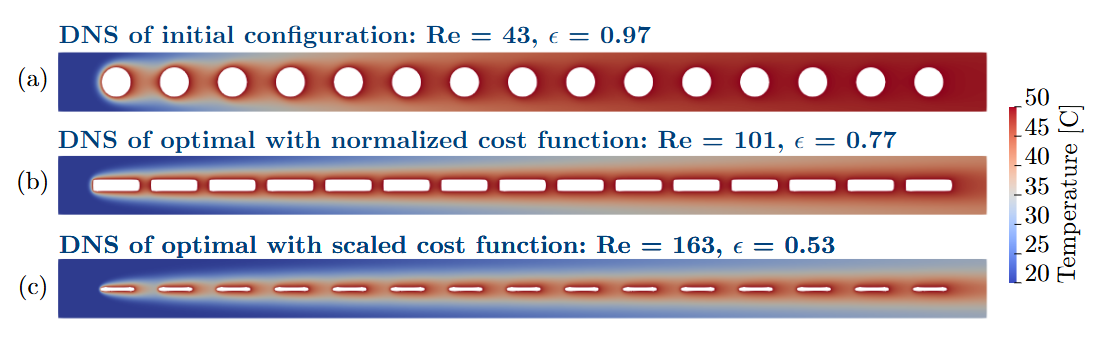}
	\caption[]{Initial (a) and optimized fin arrays obtained by optimizing the normalized (b) and scaled (c) cost function applied at unit-cell level, with the color representing the temperature $T$.
	}%
	\label{fig:dpopt}
\end{figure*}

After completing the unit-cell shape optimization, a direct numerical simulation (DNS) of the full array is performed at the initial array pressure drop, to assess the true performance increase. In \fref{fig:dpopt}, the final fin shape and temperature profile on the array are shown for both cost functionals, i.e. the normalized (\Eref{eq:cost1}) and scaled (\Eref{eq:cost2}) cost functional). 

A first observation is that the choice of cost function normalization fundamentally influences the optimization outcome. While the shape in \fref{fig:dpopt}b is only kept from forming a microchannel array by the geometric constraints in the optimization, the optimization with scaled cost function in \fref{fig:dpopt}c generates an array of slender profiles. A second observation is that both cost functions enable a significant increase in heat transfer, with an 86 \% and a 104\% increase in heat transfer for the normalized and scaled objective, respectively. 

When investigating this heat transfer increase in depth, it can be noted that it is driven by a strong increase of flow rate by a factor of 2.5-4, which is realized by a decreased drag coefficient of the fin array. It should be noted that this increase therefore comes with drawbacks. First of all, the pumping power is increased by a factor 2.3 and 3.8 for the normalized and scaled objective, respectively. Secondly, if the application is a two-phase heat exchanger and not a heat sink, a strong decrease in effectiveness is observed. Indeed, defining the heat transfer effectiveness $\epsilon$ as  
\begin{equation}
	\epsilon = \frac{{\Tout}-\Tin}{\Ts-\Tin},
\end{equation}
with ${\Tout}$ the bulk mean temperature at the outlet, a decrease from an effectiveness of 97 \% to 77 \% or 53 \%, depending on the choice of cost function. This is a clear motivation to consider different optimization problem formulations, such as the flow rate-constrained one of \Eref{eq:16}, of which the results will be discussed in \sref{sec:43}. 

\begin{figure*}
	\centering
	\includegraphics[width=0.9\textwidth]{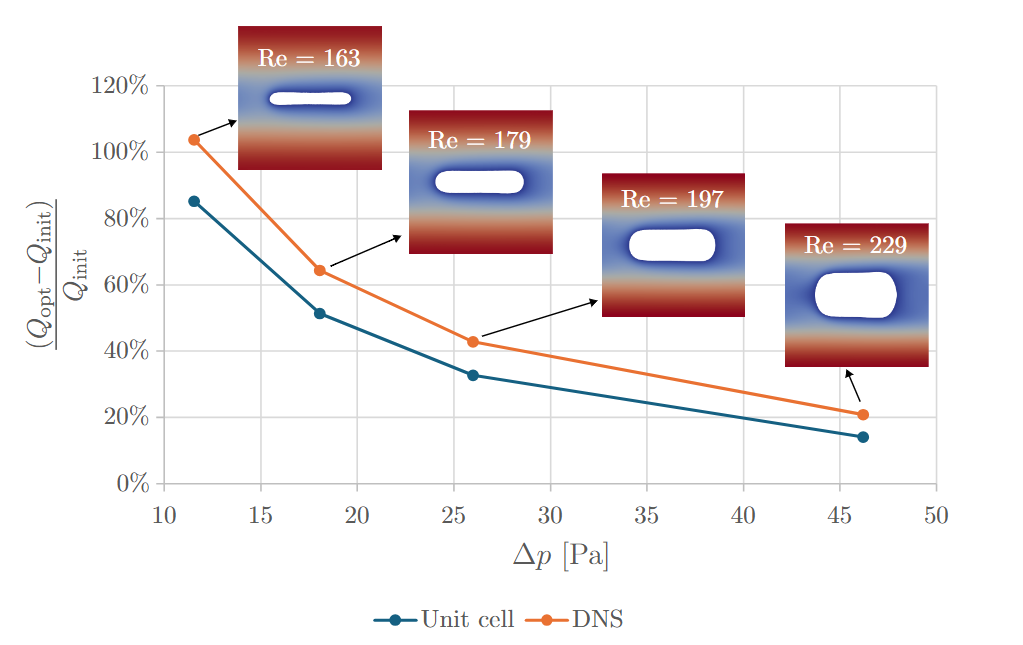}
	\caption[]{Heat transfer increase achieved by optimization as a function of pressure drop, for the cost function without $\Td$-normalization. The shapes corresponding to the optima are illustrated, with the color scale representing the periodic temperature $\theta$.
	}%
	\label{fig:pareto}
\end{figure*}

In \fref{fig:pareto}, we finally examine the potential of this unit-cell optimization approach for different imposed pressure drops. The heat transfer increase $\slfrac{\left(Q_\textrm{opt}-Q_\textrm{init}\right)}{Q_\textrm{init}}$ can be seen to decrease for increasing pressure drops and flow rates, both for the unit-cell prediction as well as for the DNS verification. Note that the subscripts ``init" and ``opt'' refer here to the heat tranfer rates of the cylinder and optimized arrays, respectively. For higher pressure drops, the optimal geometry itself clearly shows a tendency towards an increased heat transfer area rather than a reduced drag coefficient. By comparing the predicted increase from the unit-cell approach with that of the DNS simulations, one can observe an error of the unit cell approach that ranges from 20 to 40 \%.   
These errors are due to the entrance effects that are absent in the unit-cell model, as will be discussed in \sref{sec:44}. Since the actual improvement even exceeds the predictions, this is not necessarily a problem.

\subsection{Fin shape optimization for imposed pressure drop and flow rate}\label{sec:43}
As explained in \sref{sec:42}, it is interesting to investigate to what extent heat transfer increase can also be achieved without increasing the flow rate. Maximizing the heat transfer for imposed pressure drop and flow rate, boils down to increasing the heat transfer effectiveness of this fin array. Since the effectiveness of the initial pin-fin array of \fref{fig:dpopt}a was already 0.97, we choose to start the optimization for the same cylindrical pin-fin array but with a pressure drop of {46 \textrm{Pa}}, corresponding to an initial effectiveness of {$\epsilon = 0.74$}. 

\begin{figure*}
	\centering
	\includegraphics[width=0.85\textwidth]{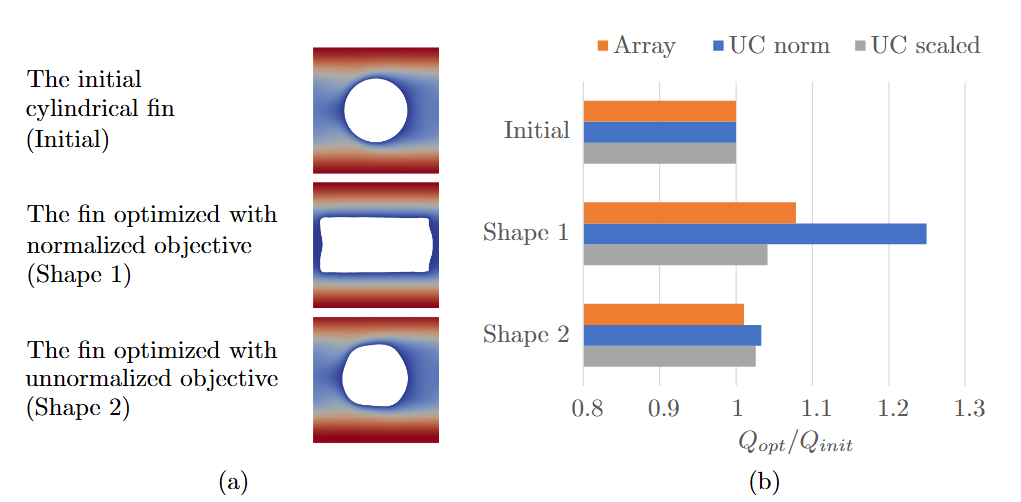}
	\caption[]{(a) Initial and optimized shapes for the optimization with imposed flow rate and pressure drop. (b) Heat transfer gains for optimal shapes 1 and 2, evaluated 
    the array simulation (orange), and normalized (blue) and scaled (gray) unit-cell heat transfer predictions.
	}%
	\label{fig:flowopt}
\end{figure*}

The optimal shapes resulting from an optimization with the normalized cost (shape 1) and scaled (shape 2) cost functional, as well as their respective heat transfer increase, are given in \fref{fig:flowopt}. 
Once again, the way the cost function is scaled yields markedly different optimized shapes.
Another observation is that the heat transfer increase observed in the DNS simulation is limited to 8 \% for shape 1, and 1 \% for shape 2. The latter improvement remains in the margin of discretization errors. While it is easily understood that the heat transfer improvement is much more limited for this heavily constrained optimization problem, there is a substantial difference with the performance increase predicted by the unit-cell approach. Especially for the optimization with normalized objective, this discrepancy is very pronounced with a predicted increase of 25 \%, compared to 8 \% in the array simulation. Therefore, the sources of this error are investigated more in depth in the next section.

\subsection{Investigation of error contributions}\label{sec:44}
The discrepancy between the unit-cell estimate of the heat transfer and the heat transfer from the array simulation can be expected to consist of two main model errors: 1) the assumption that the values of $\To$ and $\Tob$ remain fixed throughout optimization, and 2) the negligence of entrance effects.

\begin{figure*}
	\centering
	\includegraphics[width=0.9\textwidth]{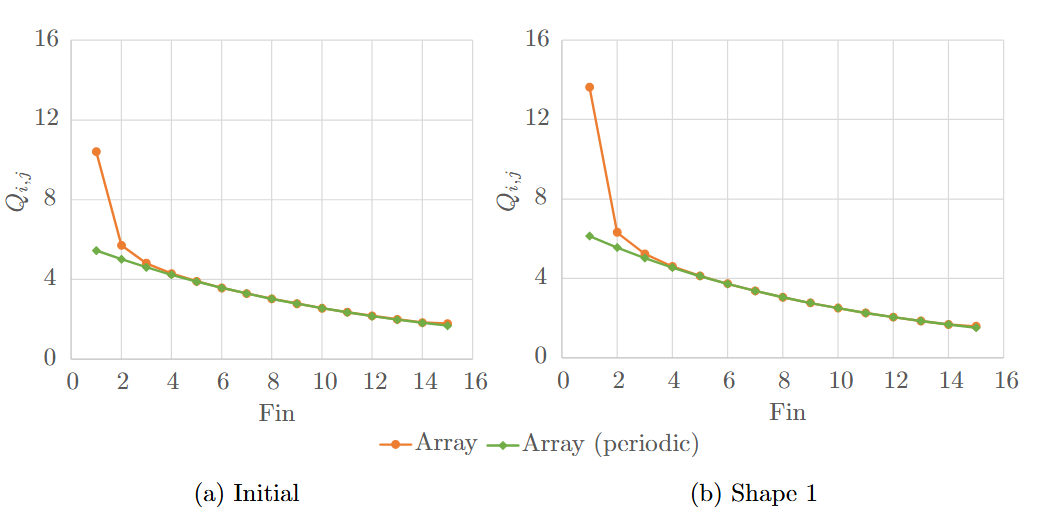}
	\caption[]{ The heat transfer $\Qij$ per fin from the array simulation {(orange)} and its periodic contribution {(green)} for the initial fin (a) and fin shape 1 (b).
	}%
	\label{fig:arrayheat}
\end{figure*}

In an attempt to quantify the error introduced by each assumption, the array heat transfer rate $\Qa$ is decomposed in a periodic contribution $\Qp$ that follows the macro-scale exponential decay and the entrance contribution $\Delta Q_\textrm{periodic} = \Qa - \Qp$. Figure \ref{fig:arrayheat} shows this decomposition for each fin separately for the initial shape and shape 1, along a single fin row. In addition, we check the error induced by assuming $\To,\Tob$ are fixed throughout the optimization. 
To this end, the actual values of $\To$ and $\Tob$ are calculated from the DNS simulations of the initial and optimized arrays. Hereto, an exponential is fitted to the periodically developed part of the volume-averaged temperature $\av{T}$ along the array. The intercept of this exponential then represents the value of $\To$ and $\Tob = \To \slfrac{\norm{\Td}_\infty}{\av{\Td}}$. From these intercepts and value of $\Td$ at the upper boundary, the values of $\To$ and $\Tob$ can be obtained.  The error of assuming $\To$ fixed is then found by taking the difference between the heat transfer $\Qtot$ estimated from the unit cell approach and the corrected heat transfer $\Qtot \slfrac{\To}{T_{0,\textrm{init}}}$, i.e. 
\begin{equation}
\Delta Q_{\To} = \Qtot \left(\slfrac{\To}{T_{0,\textrm{init}}}-1\right).    
\end{equation}
Analogously, the error from assuming $\Tob$ fixed in the scaled temperature decomposition can be retrieved as 
\begin{equation}
\Delta Q_{\Tob} = \Qtot \left(\slfrac{\Tob}{T_{1,\textrm{init}}}-1\right).    
\end{equation}
The error contributions are summarized in table \ref{tab:errors}.

It can be seen that the large error observed for the normalized heat transfer formulation is mostly due to the assumption of constant $\To$ in the temperature decomposition $T -\Ts=  \To \frac{\Td}{\av{\Td}} \exp\left(\lT x\right) $. Indeed, after optimization, the value of $\av{\Td}$ has decreased by a factor {1.21}. The optimization exploits this model deficiency to increase $\slfrac{\Td}{\av{\Td}}$ and thus the estimated inlet temperature. 
 The value of $\To$ in turn decreases by a factor of {1.20} after optimization. It therefore almost completely sets off the corresponding change in $\av{\Td}$, as expected. Thus, the error on the unit-cell heat transfer prediction becomes as high as 18 \% for the normalized heat transfer objective of shape 1. 

This clear drawback of the normalized cost function is an argument for the alternative scaled temperature decomposition, $T -\Ts=  \Tob \slfrac{\Td}{\norm{\Td}_\infty} \exp\left(\lT x\right)$. 
Indeed, as can be observed in \fref{fig:flowopt}, $\Td$ systematically reaches its maximal value at the corner when the barycenter constraint is imposed.  
Since this maximal value of $\Td = 1$ is kept fixed during the optimization, the optimization with scaled cost function cannot exploit strong local increases in $\Td$ to fictitiously increase the temperature at the inlet and reach a higher heat transfer. 
When this cost functional is used, entrance effects are clearly the dominant error contribution with errors of 10-15 \%, compared to errors coming from changes in $\Tob$ as low as 1-3 \%. 

While the agreement between the scaled cost function and the array simulation is clearly better, the effective heat transfer increase reached when optimizing with this cost function is much lower, as can be seen for shape 2 in \fref{fig:flowopt}b. It is counterintuitive that the normalized cost function representation that is observed to be the least accurate, provides the highest heat transfer increase. The explanation is that the largest share of the heat transfer increase is coming from the increase in entrance heat transfer, as can be clearly seen from \fref{fig:arrayheat}. It can therefore be concluded that this increase is not due to a more advantageous cost functional formulation.
 Rather, for the small array under consideration, where entrance effects significantly contribute to total heat transfer, the unit-cell model may simply be insufficient. 
Future research should therefore consider extending the model to account for entrance effects. Nevertheless, it should be pointed out that the proposed methodology is very attractive with a reduction in mesh size by a factor that is roughly equal to the amount of fins in a single row (here 15). 


\begin{table*}
    \centering
    \resizebox{\textwidth}{!}{ 
    \renewcommand{\arraystretch}{1.2}
    \begin{tabular}{|c|ccc|cccc|ccc|}
        \hline
        \multirow{2}{*}{\textbf{Geometry} \rule{0pt}{5ex}} & \multicolumn{3}{c|}{\textbf{Array} \rule{0pt}{5ex}} & \multicolumn{4}{c|}{\textbf{Unit cell (normalized)} \rule{0pt}{5ex}} & \multicolumn{3}{c|}{\textbf{Unit cell (scaled)} \rule{0pt}{5ex}} \\ 
        & $\Qa \left[\frac{\textrm{W}}{\textrm{m}}\right]$ & $\Qp \left[\frac{\textrm{W}}{\textrm{m}}\right]$ & $\left| \frac{\Delta Q_{entrance}}{\Qa} \right|$ & $\Qtot \left[\frac{\textrm{W}}{\textrm{m}}\right]$ & $\Qtot \frac{\To}{T_{0,\textrm{init}}} \left[\frac{\textrm{W}}{\textrm{m}}\right]$ & $\left| \frac{\Delta Q_{T_0}}{Q_{array}} \right|$ &  & $\Qtotb \left[\frac{\textrm{W}}{\textrm{m}}\right]$ & $\Qtotb \frac{\Tob}{T_{1,\textrm{init}}} \left[\frac{\textrm{W}}{\textrm{m}}\right]$ & $\left| \frac{\Delta Q_{T_1}}{Q_{array}} \right|$ \rule{0pt}{2.5ex} \\
        \hline
        Initial & 54.32 & 48.44 & 10.8\% & 51.73 & 51.73 & 0.0\% & & 51.73 & 51.73 & 0.0\% \\
        Shape 1 & 58.79 & 50.16 & 14.7\% & 64.61 & 53.93 & 18.2\% & & 53.85 & 54.48 & 1.1\% \\
        Shape 2 & 55.07 & 48.63 & 11.7\% & 53.44 & 52.01 & 2.6\% & & 53.04 & 51.91 & 2.1\% \\
        \hline
    \end{tabular}
    }
    \caption[]{Comparison of the heat flux estimates from the scaled and normalized unit-cell models, w.r.t to the array reference value. Moreover, $\Delta Q_\textrm{entrance}$ and $\Delta Q_{T_{0,1}}$ are estimates of the expected unit-cell error coming from the entrance effects and the assumption of constant $\To$ and $\Tob$, respectively.	}%
    \label{tab:errors}
\end{table*}



\section{Conclusions and outlook}

In this study, we presented a unit-cell shape optimization approach aimed at maximizing heat transfer in periodic fin arrays under constant solid temperature conditions. By leveraging recently developed macro-scale models for isothermal solid structures and combining them with a geometrically-constrained free-shape optimization method, we successfully obtained optimal fin geometries that maintain minimal fin distance. The computational cost of evaluating the heat transfer of the complete fin array is hereby reduced to that of a single unit cell, making our approach computationally efficient and practical for large-scale applications. Our results demonstrated significant heat transfer improvements, with increases up to 104 \% in a fictitious use-case. However, these increases were mainly achieved by a corresponding increase in flow rate. When flow rate constraints were introduced to maintain high effectiveness, the heat transfer increase was limited to 8 \%.


Our case study revealed that the scaling of the objective function significantly affects the optimization outcome and modeling accuracy.
Increased errors in heat transfer prediction arose with the normalized cost functional formulation due to the assumption of a fixed scaling factor in macroscale temperature variation, artificially elevating inlet temperature and heat transfer. In contrast, a scaled unit-cell formulation with a fixed maximum periodic temperature contribution significantly improved accuracy, reducing induced by assuming a fixed scaling factor to 1–3\%. In addition, there are increased errors in the entrance region, accounting for 10-15 \% of the total heat transfer for the investigated fin arrays. Although the discrepancy between the heat transfer increase in the unit cell approach and the full array simulation is reduced by the scaled objective, other objective and optimization problem formulations are possible. Investigating other formulations and their performance for a wider range of flow conditions and array lengths is left for future work. In particular, improved models to account for entrance effects in unit-cell models should be investigated.

Overall, our approach provides a promising method for optimizing heat transfer in periodic fin arrays, offering substantial improvements while strongly increasing computational efficiency. Future work will focus on addressing the identified limitations and exploring the potential of this method in practical applications.

\section*{Statements and declariations}
\subsection*{Funding}
Maarten Blommaert and Geert Buckinx have received funding from Research Foundation - Flanders (FWO) and the Flemish Institute for Technological Research (VITO) under FWO-VITO postdoctoral grants for parts of this work. Arthur Vangeffelen and Mehmet Basaran were funded for this research by VLAIO under project number HBC.2021.0801/IAMHEX. Mehmet Basaran also received partial funding from a KU Leuven starting grant with reference STG/21/016.
\subsection*{Conflict of interest statement}
The authors have no relevant financial or non-financial interests to declare.
\subsection*{CRediT author statement}
\textbf{Maarten Blommaert:} conceptualization, methodology, software, investigation, formal analysis, writing - original draft, funding acquisition
\textbf{Arthur Vangeffelen:} methodology, software, investigation, formal analysis, validation, writing - review \& editing
\textbf{Mehmet Basaran:} investigation, formal analysis, writing - review \& editing, visualisation
\textbf{Geert Buckinx:} conceptualization, methodology, software, writing - review \& editing
\textbf{Martine Baelmans:} conceptualization, writing - review \& editing, supervision, funding acquisition
\subsection*{Data availability}
The data sets produced in this paper are available from the corresponding author upon reasonable request.
\subsection*{Ethics approval and Consent to participate}
Not applicable
\subsection*{Replication of Results}
The paper contains all information necessary to reproduce the results. In case of further queries, please contact the corresponding author.

\begin{appendices}
\end{appendices}


\bibliography{./library,heattransfer,other}

\end{document}